\documentclass[aps,prb,twocolumn,showpacs,superscriptaddress]{revtex4}

\usepackage[pdftex]{graphicx}
\usepackage[pdftex]{epsfig}
\usepackage{epstopdf}
\usepackage{color}
\usepackage{amssymb,amsmath}
\usepackage{graphicx}
\usepackage{dcolumn}
\usepackage{multirow}

\begin{document}Ê

\title{Two-electron dephasing in single Si and GaAs quantum dots }
\author{John King Gamble}
\affiliation{Department of Physics, University of Wisconsin-Madison, Madison, WI 53706, USA}
\author{Mark Friesen}
\author{S. N.~Coppersmith}
\affiliation{Department of Physics, University of Wisconsin-Madison, Madison, WI 53706, USA}
\author{Xuedong Hu}
\affiliation{Department of Physics, University at Buffalo, SUNY, Buffalo, NY 14260}

\pacs{73.21.La,03.65.Yz,71.38.-k,85.35.Be,42.50.Lc}Ê

\begin{abstract}
We study the dephasing of two-electron states in a single quantum dot in both GaAs and Si.
We investigate dephasing induced by electron-phonon coupling and by charge noise analytically for pure orbital excitations in GaAs and Si, as well as for pure valley excitations in Si.
In GaAs, polar optical phonons give rise to the most important contribution, leading to a typical dephasing rate of $\sim 5.9$~GHz.
For Si, intervalley optical phonons lead to a typical dephasing rate of $\sim 140$~kHz for orbital excitations and $\sim 1.1$~MHz for valley excitations.
For harmonic, disorder-free quantum dots, charge noise is highly suppressed for both orbital and valley excitations, since neither has an appreciable dipole moment to couple to electric field variations from charge fluctuators.
However, both anharmonicity and disorder break the symmetry of the system, which can lead to  increased dipole moments and therefore faster dephasing rates.
\end{abstract}

\maketitle

\section{Introduction}
Both Si and GaAs quantum dot technologies are now well-established as candidates for the implementation of scalable quantum computation.\cite{Nowack:2011p1269,Maune:2012p344}
Working qubits have been fabricated with several different architectures involving singly-occupied dots, including the single-spin qubit architecture,\cite{Loss:1998p120,Morello:2010p687,Elzerman:2004p431,Koppens:2006p766,Kouwenhoven:2001p701,Ciorga:2000p16315} the  singlet-triplet scheme,\cite{Levy:2002p1446,Petta:2005p2180,Petta:2005p161301,Prance:2012p046808,Maune:2012p344} and three-dot logical qubits.\cite{DiVincenzo:2000p1642,Gaudreau2012p54} 

Recently, we proposed a new ``hybrid" qubit architecture\cite{Shi:2011unpub2} with three electrons in two quantum dots that is potentially advantageous because fast qubit operations can be performed in a relatively simple architecture. 
The hybrid design is capable of all-electrical, fast qubit gates, but the qubit wavefunctions have some charge character that gives rise to decoherence mechanisms that are not present in pure spin qubits.
This means that characterizing decoherence becomes a most pressing issue, because charge decoherence is typically much faster than spin decoherence.

In the singlet-triplet qubit, when there is a finite exchange-induced energy splitting between the singlet and triplet states, the dominant sources of decoherence have been found to be charge noise\cite{Coish:2005p125337,Hu:2006p100501,Culcer:2009p073102} and electron-phonon coupling.\cite{Hu:2011p165322} 
Much of the physics of the hybrid qubit is similar to that of the singlet-triplet qubit, with the main differences in the decoherence properties arising because one of the dots contains two electrons.
In the doubly-occupied dot, both electrons in a singlet can occupy the orbital ground state, while at least one of the triplet electrons must lie in an excited state. 
A further complicating factor in silicon-based quantum dot devices is the presence of two nearly degenerate, low-lying valley states.\cite{Boykin:2004p115,Boykin:2004p165325,Culcer:2009p205302,Friesen:2006p202106,Friesen:2007p115318,Friesen:2010p115324,Saraiva:2009p081305,Saraiva:2011p155320} These levels are split near a sharp interface by an amount that is typically comparable to the orbital energy spacing.
Hence, single-electron first excited states have two characteristic types: orbital, where the electron occupies the same valley state but the P-like first excited state of the lateral confinement potential, and valley, where the electron is in the orbital ground state and a higher valley state. 

In this paper, we calculate singlet-triplet dephasing rates in a doubly-occupied quantum dot, extending the spin relaxation calculations previously done for both GaAs\cite{Golovach:2001p045328} and Si.\cite{Prada:2008p1187} 
Unlike relaxation, pure dephasing is due to processes that conserve spin and does not involve energy exchange with the environment. 
For both electron-phonon coupling and charge noise, we consider the limiting cases of purely orbital (for both GaAs and Si) and purely  valley (for Si) excited states. 

We find that for GaAs, polar optical phonons are the main source of dephasing, leading to a decoherence rate of $\sim 5.9$~GHz. For Si, the phonon-mediated dephasing rate depends on the type of excitations supported by the quantum dot. For a first excited state that is orbital-like, intervalley optical phonons lead to a decoherence rate of $\sim 140$~kHz. For a valley-like first excited state, this same phonon channel results in a faster decoherence rate of $\sim 1.1$~MHz. For a perfectly harmonic, disorder-free dot, we find that dephasing due to charge noise is strongly suppressed. This is because the effective dipole moment between the singlet and triplet states vanishes for both orbital- and valley-like excitations. If we allow for anharmonicity and an effective dipole moment, we find that both phonon and charge noise dephasing channels are of similar strengths in Si, but phonons are the limiting mechanism in GaAs. Assuming a gate operation speed of 10 GHz (quite feasible for the hybrid qubit), the decoherence rate in silicon is consistent with the achievement of $10^4$ operations per coherence time, while the decoherence rate in GaAs is too fast for viable hybrid qubit operation.

This paper is organized as follows. In Sec.~\ref{two_electron_states}, we briefly review the quantum states that are relevant to the system considered in this paper. Next, in Sec.~\ref{electron_phonon_dephasing} we formulate the problem of intra-dot singlet-triplet dephasing due to the electron-phonon coupling, following the formalism of Ref.~\onlinecite{Hu:2011p165322}. We consider first GaAs, then both pure orbital and valley excitations in Si. In Sec.~\ref{charge_noise}, we calculate dephasing due to charge noise, and compare to phonon-induced dephasing. Finally, in Sec.~\ref{discussion} we discuss the role that the dephasing mechanisms we have addressed are likely to play for qubits and suggest methods for mitigating decoherence.

\section{Two-electron states in a quantum dot}\label{two_electron_states}
The lowest energy eigenstates of two electrons in a single quantum dot (neglecting spin-orbit interaction)  are the singlet
\begin{equation}\label{singlet_def}
\left| S \right> = \left| \psi_0 \right> \frac{\left| \uparrow \downarrow \right> - \left| \downarrow \uparrow \right>}{\sqrt{2}}
\end{equation}
and the three triplet states
\begin{equation}\label{triplet_def}
\left| T \right> = \left| \psi_1 \right> \times
\begin{cases}
\left| \uparrow \uparrow \right> & (T_+)\\
\left( \left| \uparrow \downarrow \right> + \left| \downarrow \uparrow \right>\right)/\sqrt{2} & (T_0)\\
\left| \downarrow \downarrow \right> & (T_-)
\end{cases},
\end{equation}
where $\left| \psi_0 \right>$ is the two-electron spatial ground state and $\left| \psi_1 \right>$ is the first excited state.
We work within the Heitler-London approximation;\cite{Hu:2011p165322}
in this approximation, $\left| \psi_0 \right> \approx \left| 0 0 \right>$, indicating that both electrons are in their single-electron ground states, and $\left| \psi_1 \right> \approx \left( \left| 0 1 \right> - \left| 1 0 \right> \right)/\sqrt{2}$, where one electron is in its ground state and the other is in its first excited state. 
This paper considers dephasing due to the electron-phonon coupling and charge noise, both of which conserve spin, so from here on we will focus on the spatial component, with the appropriate spin wavefunction understood. We refer to the triplets collectively as $\left| T \right>$ when the particular spin configuration is not important.

In GaAs, the wavefunctions of the ground and first excited states have identical dependence on $z$ (the direction perpendicular to the quantum well), but have S- and P-like transverse envelopes in the x-y plane (the plane of the quantum well). For a quadratic quantum dot confinement potential, the effective mass approximation for the ground state wavefunction is
\begin{equation}\label{GaAsGroundState}
\psi_0^{GaAs}(\mathbf r) = u(\mathbf r) \frac{F(z)}{\sqrt{  \pi} L}e^{-(x^2+y^2)/(2 L^2)}
\end{equation}
while the first excited state wavefunction is
\begin{equation}\label{GaAsFirstState}
\psi_1^{GaAs}(\mathbf r) = u(\mathbf r)\frac{F(z)(x+iy)}{\sqrt{ \pi} L^2}e^{-(x^2+y^2)/(2 L^2)},
\end{equation}
where $u(\mathbf r)$ is the periodic (with the lattice periodicity) component of the Bloch function at the conduction band minimum (the $\Gamma$-point in GaAs), $F(z)$ is the envelope function along z, and $L$ is the lateral extent of the wavefunction. 

For silicon, we write the ground state wavefunction as 
\begin{equation}\label{SiOrbitalGround}
\psi_0^{Si}(\mathbf r) =\phi_-(\mathbf r) \frac{F(z)}{\sqrt{ \pi} L} e^{-\frac{1}{2 L^2}(x^2+y^2)},
\end{equation}
where
\begin{equation}
\phi_\pm(\mathbf r ) =  \frac{u_+(\mathbf r) e^{i k_0 z} \pm u_-(\mathbf r) e^{- i k_0 z} }{\sqrt{2}},
\end{equation}
and $u_\pm(\mathbf r)$ is the periodic part of the Bloch function evaluated at the conduction band minimum located at $\pm k_0 \hat z$. In a quantum dot fabricated in a strained silicon quantum well,  $k_0 \approx 0.82 \times 2 \pi/a$, with $a=0.543$ nm the length of the Si cubic unit cell.\cite{Friesen:2007p115318}
Depending on the magnitude of the valley splitting introduced by the sharp interfaces and the electric field in the z-direction, the lowest energy excited states can either be valley-like or orbital-like. For the case of an orbital-like excitation (large valley splitting), the wavefunction is
\begin{equation}\label{SiOrbitalFirst}
\psi_1^{Si,O}(\mathbf r) = \phi_-(\mathbf r) \frac{F(z)(x+iy)}{\sqrt{ \pi} L^2}e^{-\frac{1}{2 L^2}(x^2+y^2)}.
\end{equation}

When the valley splitting is smaller than the orbital splitting, both the ground state and first excited state have S-like transverse wavefunctions, but their z-direction wavefunctions are different valley states. The ground state wavefunction is still given by Eq.~(\ref{SiOrbitalGround}), but the first excited state is now
\begin{equation}\label{SiValleyFirst}
\psi_1^{Si,V}(\mathbf r) = \phi_+(\mathbf r) \frac{F(z)}{\sqrt{ \pi} L} e^{-\frac{1}{2 L^2}(x^2+y^2)}.
\end{equation}

The periodic parts of the Bloch functions, $u(\mathbf r)$, have discrete Fourier spectra, with contributions occurring at reciprocal lattice vectors $\mathbf G$. Hence, when performing calculations with the full wavefunctions, as we do in this paper, it is convenient to decompose $u(\mathbf r)$ into the sum:\cite{Wood:1996p7949,Saraiva:2009p081305,Saraiva:2011p155320}
\begin{equation}
u(\mathbf r) = \sum_{\mathbf G} \alpha(\mathbf G) e^{i \mathbf r \cdot \mathbf G},
\end{equation}
where the expansion coefficients $\alpha(\mathbf G)$ are independent of $\mathbf r$. 
For low-frequency dephasing channels such as acoustic phonons and charge noise only the $\mathbf G = \mathbf 0$ mode contributes significantly.\cite{Mahan:2000}
For high-frequency processes such as optical phonon couplings, contributions with $\mathbf G \neq \mathbf 0$ can also be important.
We assume that the optical electron-phonon couplings are independent of $\mathbf G$. 
With this assumption, one can prove that the calculation of dephasing rates is independent of the form of the periodic part of the Bloch functions.
Hence, for all instances we consider in this paper, we may ignore the periodic part of the Bloch functions.

In a real system, disorder would cause the excited states to have mixed valley and orbital characteristics,\cite{shi:2011p233108} for which pure orbital-like and pure valley-like first excited states represent limiting cases. Hence, it is important to consider both the pure valley and orbital excitations described above. 

\section{Dephasing via the electron-phonon interaction}\label{electron_phonon_dephasing}
We now consider the dephasing of two-electron states in a single quantum dot due to the electron-phonon interaction, following the techniques of Ref.~\onlinecite{Hu:2011p165322}. In Sec.~\ref{phonons_GaAs}, we consider dephasing in GaAs due to deformation potential, longitudinal and transverse piezoelectric, and polar optical phonons. In Sec.~\ref{phonons_Si},
we turn to silicon, where we may have either valley or orbital excitations, and the relevant dephasing channels are through intravalley deformation potential and intervalley optical phonons.

The general form of the electron-phonon interaction is \cite{Mahan:2000}
\begin{equation}
V_{ep}(\mathbf r) = \sum_{\mathbf q, \mathbf G, \lambda} M_\lambda(\mathbf q + \mathbf G) \rho(\mathbf q + \mathbf G) (a_{\mathbf q,\lambda}+ a_{-\mathbf q,\lambda}^\dagger),
\end{equation}
where each $\mathbf G$ is a reciprocal lattice vector, $\mathbf q$ is constrained to the first Brillouin zone, $\rho$ is the electron density operator, $a_{\mathbf q,\lambda}$ and $a_{\mathbf q,\lambda}^\dagger$ annihilate and create, respectively, phonons with wave vector $\mathbf q$, and $\lambda$ indexes the phonon mode. Since we will treat the the relevant modes separately, we will suppress this sum over phonon modes in the calculations that follow.
The electron-phonon coupling, $M_{\mathbf q + \mathbf G}$, is defined by
\begin{equation}\label{jkg_M_def}
M(\mathbf q + \mathbf G) = - V_{ei}(\mathbf q + \mathbf G)\left[ (\mathbf q + \mathbf G) \cdot \mbox{ \boldmath $\xi$} \right] \sqrt{\frac{\hbar}{2 \rho_m \Omega \omega_{\mathbf q}} },
\end{equation}
where $V_{ei}$ is the electron-ion potential, {\boldmath $\mathbf \xi$} is the phonon polarization vector, $\rho_m$ is the crystal mass density per unit volume, $\Omega$ is the crystal volume, and 
$\omega_{\mathbf q}$ is the phonon frequency. 
Since the singlet and triplet have different charge distributions, they are dressed differently by the phonons.
The phonons can themselves decohere, which in turn causes dephasing between the singlet and triplet states. 
Following Ref.~\onlinecite{Hu:2011p165322}, the singlet-triplet dephasing rate due to the electron-phonon coupling is
\begin{equation}\label{jkg_ST_relax}
\Gamma_{ST} = \frac{\Omega}{2 \pi^3 \hbar^2} \sum_{\mathbf G} \int d^3 \mathbf q \frac{ \left| M(\mathbf q + \mathbf G) A(\mathbf q + \mathbf G) \right|^2}{\omega_{\mathbf q}^2 + (\gamma_{\mathbf q}/2)^2}\frac{\gamma_{\mathbf q}}{2},
\end{equation}
where $\gamma_{\mathbf q}$ is the phonon relaxation rate, and $A$ is a Fourier component of the charge density difference between triplet and singlet states:
\begin{equation}\label{def_of_A}
A(\mathbf q + \mathbf G) = \frac{1}{2} \left( \left< T \right| \rho(\mathbf q + \mathbf G ) \left| T \right> -\left< S \right| \rho(\mathbf q + \mathbf G ) \left| S \right> \right).
\end{equation}
Using the approximate forms for the singlet and triplet states detailed in Eqs.~(\ref{singlet_def}) and (\ref{triplet_def}), we have
\begin{equation}\label{jkg_A_def}
A(\mathbf q + \mathbf G) = \frac{1}{2} \left( \left< 1 \right| e^{ i (\mathbf q + \mathbf G) \cdot \mathbf r } \left| 1 \right> -\left< 0 \right|e^{ i (\mathbf q + \mathbf G) \cdot \mathbf r }  \left| 0 \right> \right),
\end{equation}
where $\left| 1 \right>$ is the single-particle first excited state and $\left| 0 \right>$ is the single-particle ground state. In the following subsections, we evaluate Eq.~(\ref{jkg_ST_relax}) for the different types of phonons in both GaAs and Si.


\subsection{Phonon-induced dephasing in GaAs}\label{phonons_GaAs}

 \begin{figure}[tb!] 
\begin{center} 
\includegraphics[width=1.0 \linewidth]{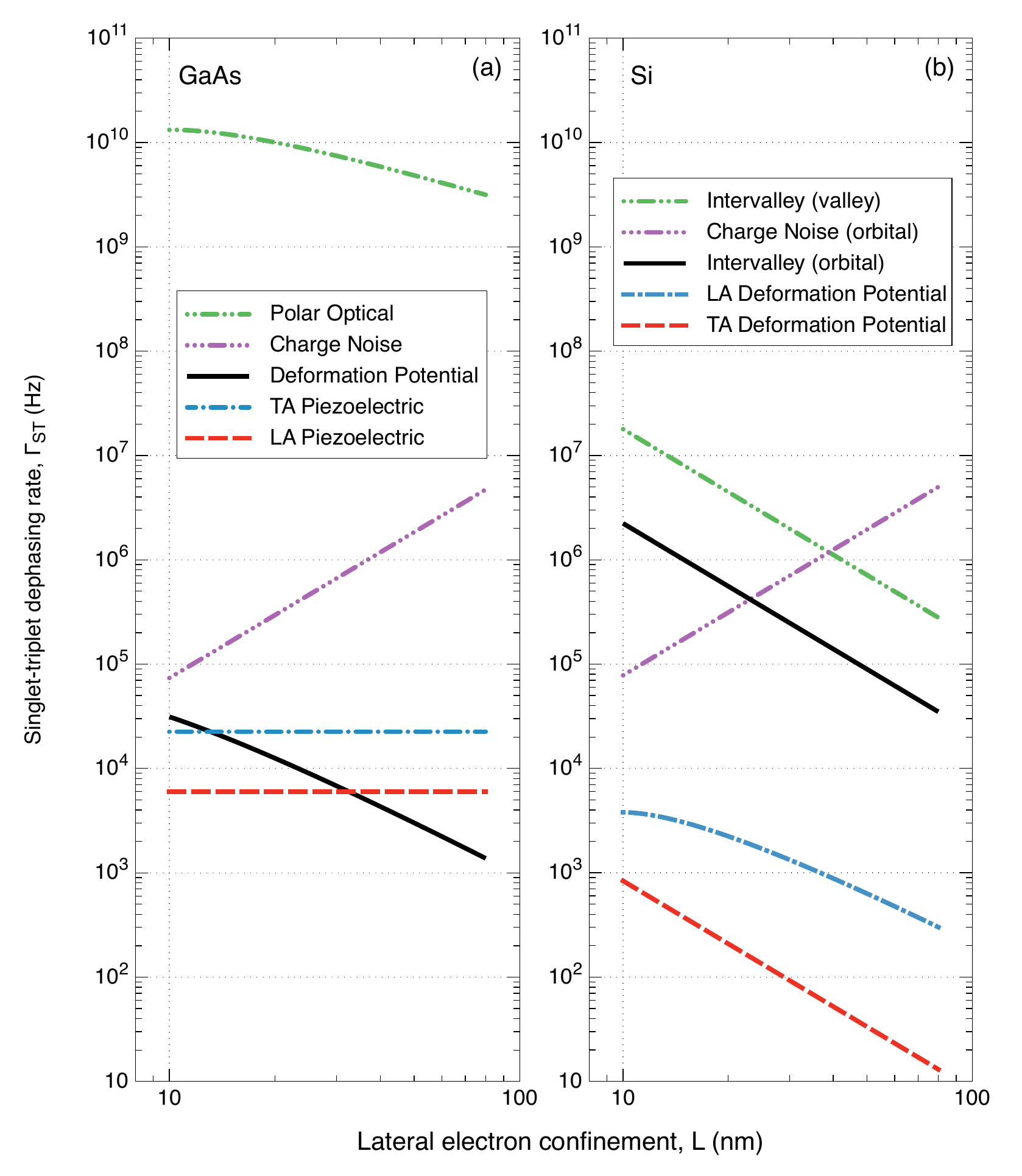}
\end{center} 
\caption{(Color online) Calculated singlet-triplet dephasing rates of two electrons in a single dot via the phonon mechanisms considered in Sec.~\ref{electron_phonon_dephasing} and the charge noise mechanism considered in Sec.~\ref{charge_noise}. (a): Plot of dephasing rates versus the lateral extent of the electron wavefunctions for the mechanisms pertinent for GaAs: polar optical phonons (Eq.~(\ref{GaAs_PO})), deformation potential phonons (Eq.~(\ref{GaAs_DP})), transverse piezoelectric phonons (Eq.~(\ref{GaAs_TA_Piezo})), and longitudinal piezoelectric phonons (Eq.~(\ref{GaAs_LA_Piezo})). 
(b): Plot of singlet triplet dephasing rates versus lateral wavefunction extent for different mechanisms in Si: intervalley phonons for valley excitations (Eq.~(\ref{Si_Valley_LO})), and intervalley phonons (Eq.~(\ref{Si_Orbital_LO})), longitudinal acoustic phonons (Eq.~(\ref{Si_LA})), and transverse acoustic phonons (Eq.~(\ref{Si_TA})) for orbital excitations. The dephasing rates due to charge noise plotted for both materials systems are determined from Eq.~(\ref{t2_conversion}), using the energy fluctuations for orbital excitations in a single dot (Eq.~(\ref{quadrupole_term_SP})), and assuming zero dipole moment.
\label{phonon_figure}
}
\end{figure}

For the purely orbital excitations supported by GaAs, the ground state transverse wavefunction is an S-orbital, while the first excited state transverse wavefunction is a P-orbital, as given in Eqs.~(\ref{GaAsGroundState}) and (\ref{GaAsFirstState}). The three types of phonon couplings that contribute to decoherence are: deformation potential, piezoelectric, and polar optical. These differ only in the form of the electron-phonon coupling $M$, so their calculations proceed similarly.

In all cases, we assume a Gaussian form for the wavefunction in the z-direction:
\begin{equation}
F(z) = \frac{1}{\pi^{1/4} \sqrt{d}} e^{-z^2/\left(2 d^2\right)},
\end{equation}
where $d$ is the confinement length along the growth axis, which is typically a few nanometers. 
Choosing this form for the wavefunction represents an approximation, but it captures the relevant physics and allows us to obtain analytic results.
Using this approximation and Eqs.~(\ref{GaAsGroundState}) and (\ref{GaAsFirstState}), we use Eq.~(\ref{jkg_A_def}) to obtain
\begin{equation}\label{AGaAs}
\left| A^{GaAs}(\mathbf q) \right|^2 = \frac{L^4 \left( q_x^2 + q_y^2\right)^2}{4}
  e^{ - \left( \left( q_x^2 + q_y^2\right) L^2 + q_z^2 d^2 \right) /2 }.
\end{equation}
Here, we have taken $\mathbf G = \mathbf 0$, since $\left| A \right|^2$ goes to zero rapidly for  $\left| \mathbf q \right| = q \gtrsim 1/d,1/L$, both of which are much smaller than the size of the first Brillouin zone. 

We first consider deformation potential phonons, for which the electron-phonon coupling $M$ has the form\cite{Hu:2011p165322}
\begin{equation}
M^{DP}_{GaAs}(\mathbf q) = D q \left( \frac{\hbar}{2 \rho_m \Omega \omega_{ q}}\right)^{1/2},
\end{equation}
where $D = 8.6$~eV is the deformation potential constant and $\rho_m=5.33\times10^3$~kg/m$^3$ is the mass density of GaAs. The angular frequency $\omega_{\mathbf q}$ is given by the standard relationship for acoustic phonons: $\omega_{\mathbf q} = v_s q$, where  $v_s = 5.2\times10^3$~m/s is the longitudinal speed of sound in GaAs, averaged over direction.\cite{Lundstrom:2000} 

The last piece of information we need to compute the dephasing rate is the phonon relaxation rate $\gamma_q$. 
Eq.~(\ref{AGaAs}) implies that $|A|^2$ is strongly peaked, so it determines the $\mathbf q$-values that contribute to the integral in Eq.~(\ref{jkg_ST_relax}).
In GaAs, $| A|^2$ goes to zero both as $q$ goes to zero and when $q \gg 1/d,1/L$. Hence, the low- and high-frequency behaviors of $\gamma_q$ are not important.
For sufficiently high frequencies at low temperature, it is expected that two frequency-dependent phonon attenuation channels will become relevant: anharmonic decay and isotope scattering.\cite{Maris:2011p024301} However, the frequencies we consider here are low enough that these mechanisms are unimportant, and the dominant source of phonon relaxation is due to interface scattering.\cite{Maris:2011private}

To obtain an estimate for the phonon relaxation rate, we use experimental measurements of phonon attenuation due to interface scattering, which were performed at low-temperatures in Si.\cite{Hao:2001p224301} 
To convert between the two, one uses $\gamma_q = 2 \alpha_q v_s$,\cite{Daly:2009p174112} where $\alpha_q$ is attenuation and $v_s = 8.49\times10^5$ cm/s  is the speed of sound in Si along $[100]$. 
 For LA phonons in Si, low-temperature measurements have shown that at low frequencies (up to 100 GHz), phonon attenuation is roughly frequency-independent, and is about $2.5$ cm$^{-1}$ along the $[100]$ direction.\cite{Hao:2001p224301} 
This translates to a low-frequency experimental limit of $\gamma_0^{LA} = 4.25$ MHz, which will serve to give us an estimate on the phonon relaxation rate. 
Since this mechanism is due to the geometry rather than the particular material properties of Si, we will also use the above relaxation time for acoustic phonons in GaAs.

By switching to polar-cylindrical coordinates, evaluation of the integral in Eq.~(\ref{jkg_ST_relax}) is straightforward. 
We set $\omega_{\mathbf q}^2 + (\gamma_q/2)^2 \approx \omega_{\mathbf q}^2$, which is valid because the frequencies that contribute to the integral satisfy $\omega_{\mathbf q} \gg \gamma_0$. 
Electrostatically defined quantum dots typically obey $L \gg d$, so we expand the integration result to first order in $d/L$, 
obtaining an expression for $\Gamma_{ST}^{GaAs,DP}$, the singlet-triplet decoherence rate in GaAs due to deformation potential electron-phonon coupling:
\begin{equation}\label{GaAs_DP}
\Gamma_{ST}^{GaAs,DP} \approx \frac{D^2 (4 \ln(2 L/d) - 3)\gamma_0^{LA}}{16 \pi^2 L^2 v_s^3 \rho_m \hbar}
\approx \frac{6.9}{L^2} \, \mathrm{MHz \, nm^2}.
\end{equation}
Here, the dependence on $L$ and $d$ can be understood by power counting in Eq.~(\ref{jkg_ST_relax}). For the polar couplings we consider next, the $L$ and $d$ dependencies can be understood by examining the $q_z$ integral in Eq.~(\ref{jkg_ST_relax}) over the range where $\left| q_z \right| \ll \left| q_x\right|,\left| q_y \right| $, followed by power counting.

We next consider piezoelectric coupling, which contribute in both longitudinal and transverse phonon modes. In this case, the electron-phonon coupling is\cite{Hu:2011p165322}
\begin{align}
M_{GaAs}^{PE}(\mathbf q ) &= \frac{2 i e e_{14}}{q^2} \left( \frac{\hbar}{2 \rho_m \Omega \omega_{ q}}\right)^{1/2}  \\
&\times \big(q_x q_y \xi_z + q_y q_z \xi_z + q_z q_x \xi_y \big) \nonumber,
\end{align}
where $e$ is elementary charge and $e_{14} = 1.38 \times 10^9$~V/m is an elasticity tensor component. For longitudinal phonons, $\mbox{\boldmath $\mathbf \xi$} = \mathbf q/q$. Integration of Eq.~(\ref{jkg_ST_relax}), expanded to  lowest order in $d/L$, yields $\Gamma_{ST}^{GaAs,PE,LA}$, the singlet-triplet dephasing rate in GaAs due to piezoelectric coupling between electrons and longitudinal acoustic phonons:
\begin{equation}\label{GaAs_LA_Piezo}
\Gamma_{ST}^{GaAs,PE,LA} \approx \frac{3 e^2 e_{14}^2 \gamma_0^{LA}}{140 \pi^2 v_s^3 \rho_m \hbar}
\approx 6.0 \, \mathrm{kHz} .
\end{equation}
In the limit when the media is considered isotropic and homogeneous, the two transverse phonon branches are degenerate, and we can choose any two orthogonal polarizations.  One possible polarization is
\begin{equation}
\mbox{\boldmath $\mathbf \xi$} = \left[ \frac{q_y}{\sqrt{q_x^2+q_y^2}},-\frac{q_x}{\sqrt{q_x^2+q_y^2}},0\right],
\end{equation}
and any rotation of this vector about $\mathbf q$ is also a valid transverse polarization. We average over this plane before integrating over $\mathbf q$. This complicates the resulting integral, but it can still be carried out analytically, yielding $\Gamma_{ST}^{GaAs,PE,TA}$, the singlet-triplet dephasing rate in GaAs due to piezoelectric coupling between electrons and transverse acoustic phonons:
\begin{equation}\label{GaAs_TA_Piezo}
\Gamma_{ST}^{GaAs,PE,TA} \approx \frac{e^2 e_{14}^2 \gamma_0^{LA}}{70 \pi^2 v_s^3 \rho_m \hbar}
\approx 22 \, \mathrm{kHz}
\end{equation}
per transverse mode.

For polar optical phonons, the electron-phonon coupling is\cite{Hu:2011p165322}
\begin{equation}
M_{GaAs}^{PO}(\mathbf q) = \sqrt{\frac{2 \pi e^2 \hbar \omega_0}{q^2 \Omega} \left(\frac{1}{\epsilon_\infty} - \frac{1}{\epsilon_0}\right)},
\end{equation}
where $\epsilon_\infty=10.89 \epsilon_{vac}$ and $\epsilon_0=12.9 \epsilon_{vac}$ are the high and low frequency limits of the GaAs dielectric function, and $\epsilon_{vac}$ is the vacuum permittivity. The frequencies of optical phonons are essentially $q-$independent, with $\hbar \omega_0 = 36.35$~meV. Since optical phonons are much higher in frequency than acoustic phonons, they also have much shorter lifetimes, with measurements indicating $\gamma_0^{LO} \approx 160$~GHz.\cite{vonderLinde:1980p1505} The integration proceeds similarly to the acoustic cases, and the resulting singlet-triplet dephasing rate $\Gamma_{ST}^{GaAs,PO}$ in GaAs due to polar optical electron-phonon coupling is, to first order in $d/L$:
\begin{align}\label{GaAs_PO}
\Gamma_{ST}^{GaAs,PO} & \approx \frac{e^2 \left(\epsilon_0 - \epsilon_\infty\right)}{\epsilon_0 \epsilon_\infty}
\frac{(3 \pi L  - 16 d) \gamma_0^{LO} }{16 \sqrt{2 \pi} L^2 \hbar \omega_0 } \nonumber \\
& \approx \frac{240}{L} \, \mathrm{GHz \, nm} .
\end{align}

In Fig.~\ref{phonon_figure}, we plot the four dephasing rates considered in this section. Typical values and scalings are listed in Table~\ref{dephasing_table}. Polar optical phonons are the largest contribution to dephasing, exceeding the others by at least five orders of magnitude. This is mainly due to the extremely fast decay of the high-frequency, optical phonons.

\begin{table*}[tb]
\caption{Typical values for the dephasing rates of the phonon-induced channels discussed in Sec.~\ref{electron_phonon_dephasing}, assuming a lateral electron confinement of $L=40$ nm and a vertical confinement along the growth direction of $d=3$ nm. The channels depend on material properties and symmetry: polar phonon couplings are absent in Si, transverse phonons do not couple electrons via the deformation potential in GaAs,\cite{Hu:2011p165322} at low temperatures and biases GaAs has only one band minimum that participates in conduction, and valley excitations in Si are not connected by low-frequency phonons. 
The scaling column describes the primary dependence of the dephasing rate on $L$, the lateral wavefunction extent and $d$, the vertical wavefunction extent. In the scalings, ($*$) indicates a neglected logarithmic correction, while ($**$) indicates that the result applies in the limit of $d/L\rightarrow 0$.  
\label{dephasing_table}
}
 \begin{center}\begin{tabular}{llcccc}
 \hline
 \hline
 \multicolumn{2}{l}{\multirow{2}{*}{Coupling mechanism}} &  \multicolumn{3}{c}{Typical dephasing rate (Hz)}&  \multirow{2}{*}{Scaling}  \\ 
 \cline{3-5}
	&	 &  GaAs &  Si: orbital excitations & Si: valley excitations &    \\ 

 \hline 
 \multicolumn{2}{l}{LA phonons}&   &   & & \\ 
  $\, \, \, \, \, $&  Deformation potential  &  $4.3\times 10^3$ &  $8.8\times 10^2$ &- &$L^{-2*}$ \\ 
  &  Piezoelectric  &  $6.0\times 10^3$ & -& - & - \\ 
\\
 \multicolumn{2}{l}{TA phonons}&   &   &  &\\ 
  &  Deformation potential  &  - &  $5.2\times 10^1$ & -&$L^{-2}$ \\ 
  &  Piezoelectric  &  $2.2\times 10^4$ &  - &-& - \\ 
\\
 \multicolumn{2}{l}{LO phonons}&   &   & & \\ 
  &  Polar optical  &  $5.9\times 10^9$ &  - & -&$L^{-1**}$ \\ 
  &  Intervalley  & - &  $1.4\times 10^5$ & $1.1\times 10^6$  &$L^{-2}d^{-1}$ \\ 
 \hline 
 \hline
 \end{tabular}\end{center} 
 \end{table*}

\subsection{Phonon-induced dephasing in silicon}\label{phonons_Si}

 \begin{figure}[tb!] 
\begin{center} 
\includegraphics[width=1.0 \linewidth]{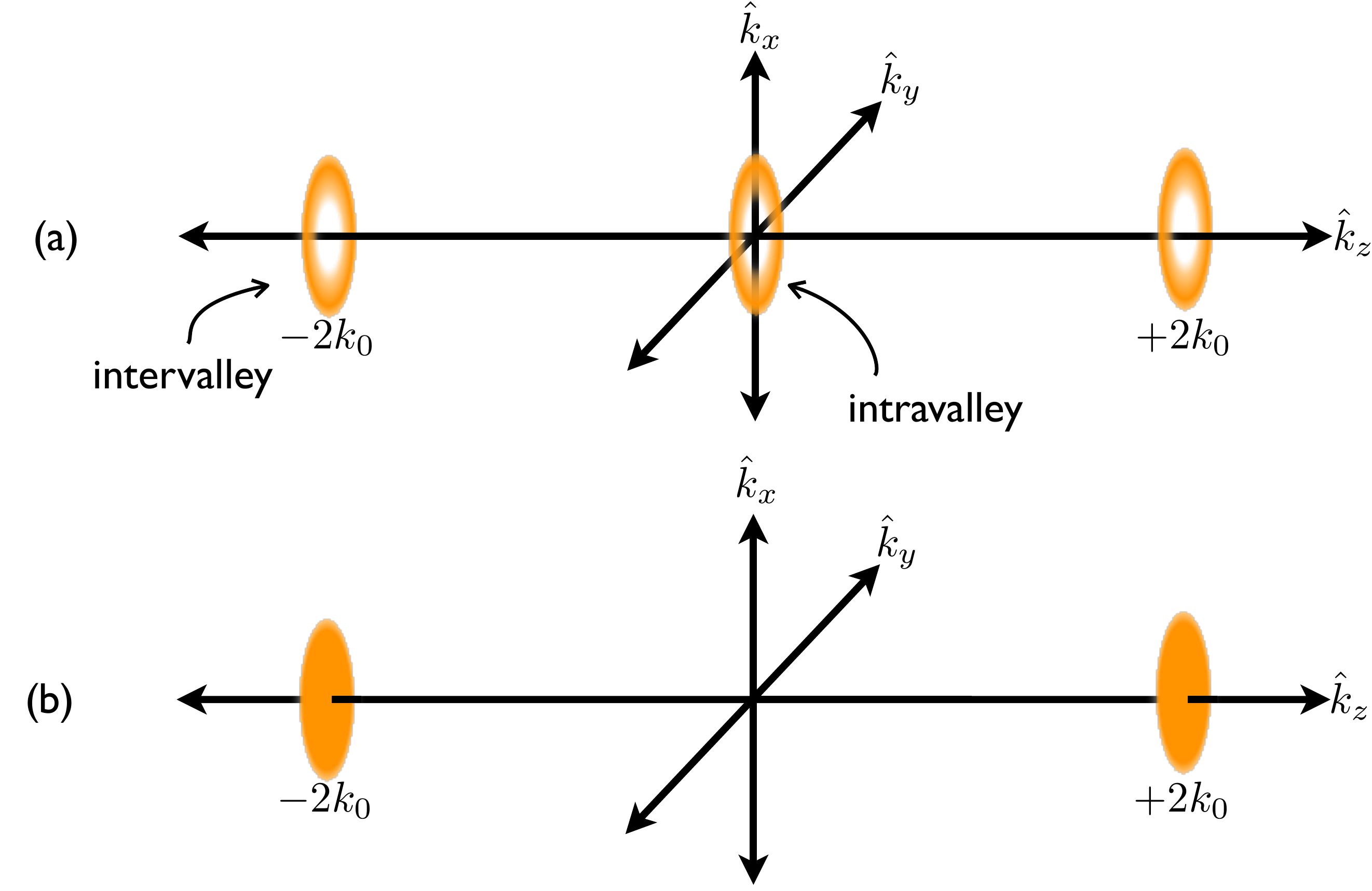}
\end{center} 
\caption{(Color online) Cartoon of the absolute value of the Fourier transform of the difference between the triplet and singlet charge distributions, $\left| A \right|^2$ (Eq.~(\ref{def_of_A})) in Si. 
Here, the shaded regions indicate the $\mathbf k-$values that contribute significantly to dephasing. 
Phonons that couple electrons in the same valley (intravalley processes) lie near the origin, while phonons that couple electrons
in different valleys (intervalley processes) lie near $\mathbf k = \pm 2 k_0 \hat k_z$.
(a): The contribution resulting from an orbital-like first excited state (Eq.~(\ref{A_Si_orbital})).
(b): The contribution resulting from a valley-like first excited state (Eq.~(\ref{A_Si_valley})). 
\label{plot_of_A}
}
\end{figure}

Unlike GaAs, Si quantum dots can support both valley and orbital excited electron states. 
The ground and first excited states in the case of an orbital excitation are given in Eqs.~(\ref{SiOrbitalGround}) and (\ref{SiOrbitalFirst}).
As we did for GaAs, we take the envelope in $z$ to be Gaussian with width $d$. 
Then, defining $\mathbf Q = \mathbf q + \mathbf G$, we evaluate Eq.~(\ref{jkg_A_def}) for orbital excitations in Si ($A_O^{Si}$), obtaining
\begin{align}\label{A_Si_orbital}
\left| A_O^{Si}(\mathbf Q) \right|^2 &=  \frac{ L^4 (Q_x^2 + Q_y^2)^2}{256}e^{-L^2(Q_x^2+Q_y^2)/2} \nonumber
 \Big( 4 e^{-d^2 Q_z^2/2}\\ 
 &+e^{-d^2 (2 k_0 + Q_z^2)/2}+e^{-d^2 (2 k_0 - Q_z^2)/2} \Big),
\end{align}
where we used the fact that the three Gaussians are well-separated in $Q_z$ to drop cross-terms. 
If we instead have excitations as given by Eqs.~(\ref{SiOrbitalGround}) and (\ref{SiValleyFirst}), we evaluate Eq.~(\ref{jkg_A_def}) for valley excited states in Si  ($A_V^{Si}$), obtaining
\begin{align}\label{A_Si_valley}
\left| A_V^{Si}(\mathbf Q) \right|^2 &= \frac{1}{4} e^{-L^2(Q_x^2+Q_y^2)/2} \\
& \times\Big( e^{-d^2(Q_z+ 2 k_0)^2/2}+e^{-d^2(Q_z- 2 k_0)^2/2} \Big). \nonumber
\end{align}
The expressions for $|A_O^{Si}|^2$ and $|A_V^{Si}|^2$ in Eqs.~(\ref{A_Si_orbital}) and (\ref{A_Si_valley}) above select the types of phonons that contribute significantly to dephasing through the integral in Eq.~(\ref{jkg_ST_relax}). 
For orbital excitations, both phonons that couple electrons within the same valley and across valleys contribute, and $|A_O^{Si}|^2$ contains contributions from	 three toroids, each with peak radius $\sqrt{2}/L$, situated in the $k_x-k_y$ planes centered at $Q_z = 0$ and $Q_z = \pm 2 k_0$. The phonons at the $Q_z = 0$ toroid correspond to intravalley processes, where $\mathbf G  = \mathbf 0$. The remaining two toroids at $Q_z = \pm 2 k_0$ are intervalley processes, where $\mathbf G = \mp( 4 \pi / a )\hat z$.
The regions of $\mathbf k-$space relevant to orbital excitations are shown in Fig.~\ref{plot_of_A} (a).
For valley excitations, $|A_V^{Si}|^2$ contains contributions from two ellipsoids centered at $\mathbf Q = (0,0,\pm 2 k_0)$. 
Since the valley-like first excited state has the same envelope function as the ground state, long wavelength phonons cannot contribute to the singlet-triplet dephasing. 
This is clearly illustrated by the vanishing of $|A_V^{Si}|^2$ at small $k$ (or long wavelength). 
The regions of $\mathbf k-$space relevant to valley excitations are shown in Fig.~\ref{plot_of_A} (b).

Now that we have identified the most important phonon wave vectors for the different dephasing mechanisms, we discuss which electron-phonon coupling mechanisms are most relevant. Since Si is not polar, the deformation potential electron-acoustic phonon coupling is the main contribution near the zone center ($\mathbf G = 0$, $q \ll 2 \pi/a$). This coupling connects electrons to both longitudinal acoustic phonons, with matrix element\cite{Hu:2011p165322}
\begin{equation}\label{si_LA}
\left| M_{LA}(\mathbf q) \right|^2 = \Xi_d^2\frac{\hbar q^2}{2 \rho_m \Omega \omega_{\mathbf q}} \left( 1 + \frac{\Xi_u}{\Xi_d} \frac{q_z^2}{q^2} \right)^2,
\end{equation}
and to transverse acoustic phonons, with coupling\cite{Hu:2011p165322}
\begin{equation}\label{si_LA}
\left| M_{TA}(\mathbf q) \right|^2 = \Xi_u^2\frac{\hbar \xi_z^2 q_z^2}{2 \rho_m \Omega \omega_{\mathbf q}},
\end{equation}
where $\Xi_d = 5.0$~eV and $\Xi_u=8.77$~eV are silicon deformation potentials. 
As in Sec.~\ref{phonons_GaAs} above, for these acoustic modes we take the phonon relaxation rate, believed to be due to interface scattering, to be $\gamma_0 = 4.25$~MHz.\cite{Hao:2001p224301} The intravalley piece of the orbital excitation is found by integration of Eq.~(\ref{jkg_ST_relax}), which proceeds very similarly to the GaAs deformation potential case we considered in Sec.~\ref{phonons_GaAs}. 
For deformation potential coupling between electrons and longitudinal phonons in Si, the singlet-triplet dephasing rate $\Gamma_{ST}^{Si,LA}$, to first order in $d/L$, is:
\begin{align}\label{Si_LA}
\Gamma_{ST}^{Si,LA} &\approx \frac{\gamma_0}{192 \pi^2 L^2 v_s^3 \rho_m \hbar} \Big[3(\Xi_d+\Xi_u)^2\ln\left(16 L^4/d^4\right)\nonumber \\
&-9\Xi_d^2-42\Xi_d\Xi_u-25\Xi_u^2\Big] \nonumber \\
& \approx \frac{1.4}{L^2} \, \mathrm{MHz \, nm^2} .
\end{align}
As was the case for GaAs, the dependence on $L$ and $d$ can be understood by power counting in Eq.~(\ref{jkg_ST_relax}).

We perform the same averaging procedure as was done for the transverse phonons in the previous section and obtain
$\Gamma_{ST}^{Si,TA}$, the singlet-triplet dephasing rate in Si due to deformation potential coupling between electrons and transverse acoustic phonons:
\begin{equation}\label{Si_TA}
\Gamma_{ST}^{Si,TA} \approx \frac{\Xi_u^2 \gamma_0}{96 \pi^2 L^2 v_s^3 \rho_m \hbar}
 \approx \frac{83}{L^2} \, \mathrm{kHz \, nm^2} .
\end{equation}

We next consider the intervalley contributions, which occur at $k_z = \pm 2 k_0$,
outside the first Brillouin zone. The reciprocal lattice vectors that contribute significantly to the relevant integrals are $\mathbf G = \pm (4 \pi/a) \hat k_z$, which give
$q_z \approx \mp 0.36 (2 \pi/a) = \mp 4.17 \times 10^9$~m$^{-1}$. 
The phonons that are responsible for this transition in silicon are due to g-type Umklapp processes.\cite{Ridley:1999}
Although symmetry restricts these to be longitudinal optical phonons, experiments indicate 
that both transverse and longitudinal acoustic phonons participate through processes in which $M(\mathbf q)$ is first-order in $\mathbf q$.\cite{Ferry:1986p1605,Ridley:1999}
However, the acoustic phonons do not play a significant role here, both because their deformation potential coupling to the electrons is weaker\cite{Lundstrom:2000} and they are much longer lived\cite{Maris:2011p024301}  than optical phonons.

The LO phonons in Si have a nearly constant energy $ \hbar \omega_0 = 62$~meV.\cite{Ridley:1999} The electron-phonon coupling arises from an optical deformation potential:\cite{Ridley:1999,Lundstrom:2000} 
\begin{equation}\label{si_LO}
\left| M(\mathbf Q)_{LO} \right|^2 = D_{if}^2\frac{\hbar}{2 \rho_m \Omega \omega_0},
\end{equation}
where the intervalley deformation potential $D_{if} = 11.0\times 10^8$ eV/cm.\cite{Lundstrom:2000} 

Finally, we estimate the relaxation rate $\gamma_q$ for optical phonons in Si. As for optical phonons in GaAs, the short-wavelength longitudinal optical phonons that cause intervalley coupling have a much shorter lifetime than the long wavelength acoustic phonons that are responsible for intravalley coupling. 
The literature value we use for the relaxation rate is $\gamma_0^{LO} = 118$ GHz. \cite{Rowlette:2008p220} For the intervalley component of the orbital excitation, we get
\begin{align}\label{Si_Orbital_LO}
\Gamma_{ST}^{Si, Orbital, LO} &= \frac{D_{if}^2 \gamma_0}{32 \sqrt{2 \pi^3} d L^2 \rho_m \hbar \omega_0^3} \nonumber \\
& \approx \frac{670}{L^2d} \, \mathrm{MHz \, nm^3} .
\end{align}
Similarly, for the case of valley excitations we have
\begin{align}\label{Si_Valley_LO}
\Gamma_{ST}^{Si, Valley, LO} &= \frac{D_{if}^2 \gamma_0}{ 4\sqrt{2 \pi^3} d L^2 \rho_m \hbar \omega_0^3} \nonumber \\
& \approx \frac{5.3}{L^2d} \, \mathrm{GHz \, nm^3} .
\end{align}

Fig.~\ref{phonon_figure} shows the dephasing rates for both GaAs and Si. Typical values and scalings with $L$ and $d$ are listed in Table~\ref{dephasing_table}. In Si, as in GaAs, most of the dephasing is due to the fast decay of optical phonons: in silicon these high-frequency phonons couple electrons across valleys. 

So far, we have only considered pure valley and pure orbital excitations. However, for non-ideal interfaces such as those with atomic steps, valley-orbit mixing occurs\cite{Friesen:2006p202106,shi:2011p233108,Friesen:2010p115324}. Since both of the limiting cases exhibit strong dephasing due to intervalley phonons (with orbital excitations suppressed by a factor of 8 from valley excitations; see Eqs.~(\ref{Si_Orbital_LO}) and~(\ref{Si_Valley_LO})), we expect that valley-orbit mixing cannot be used to suppress substantially this dephasing.

\section{Dephasing due to charge noise}\label{charge_noise}

We now consider the other expected major dephasing mechanism for our system: charge noise.\cite{Coish:2005p125337,Hu:2006p100501,Culcer:2009p073102,Ramon2010p045304} This dephasing arises because remote charge fluctuations induce random variations of the energy splitting between singlet and triplet levels by coupling to their non-equivalent charge distributions via the Coulomb interaction. These variations in the energy splitting lead to the accumulation of a  random phase between the singlet and triplet states. In turn, this introduces a phase difference between the logical qubit states. 
The dephasing mechanism and the estimated decay rates are essentially equivalent for GaAs and Si quantum dots, so we do not treat these two systems separately in this section.

Because the hybrid qubit has a potentially strong charge characteristic, it can couple to remote charge traps. 
We assume the simplest non-trivial charge fluctuation:  a single, remote charge trap with two states (occupied and empty). 
To determine the dephasing rate, the first step is to compute the effect of the change in the state of the charge trap on the singlet-triplet energy splitting. We work to first order in perturbation theory, where we may calculate the small change in energy by using the unperturbed (spatial) wavefunctions. Using the formalism of Sec.~\ref{two_electron_states}, it is straightforward to show that the first-order estimate of the variation in energy splitting $\Delta V(\tau)$ is 
\begin{align}
\Delta V(\tau)&= \left< T \right| V(\tau) \left| T \right> - \left< S \right| V(\tau) \left| S \right>  \nonumber \\
& \approx \left< 1 \right| V(\tau) \left| 1 \right> - \left< 0 \right| V(\tau) \left| 0 \right>,
\end{align}
where $\left| 0 \right>$ is the ground state, $\left| 1 \right>$ is the first excited state, and $\tau$ is time. 
Here, we assume that the energy fluctuations are much smaller than the singlet-triplet splitting, and hence also the confinement energy. 
At any instant in time, our perturbing charge trap might be occupied or empty. If the charge trap at the position $\mathbf r$ is occupied, and hence perturbing the singlet-triplet energy splitting, we have
\begin{equation}\label{deltaVdef}
\Delta V =  \int d^3 r' \frac{e^2}{4 \pi \epsilon_0} \frac{\left| \psi_1(\mathbf r' ) \right|^2-\left| \psi_0(\mathbf r' ) \right|^2}{\left| \mathbf r - \mathbf r' \right|},
\end{equation}
where $e$ is the elementary charge and $\epsilon_0$ is the (low-frequency) dielectric constant of our material. 
We assume that the trap is distant and calculate $\Delta V$ in a multipole expansion.\cite{Jackson:1999} For S- and P-like orbitals in a perfectly harmonic dot, the lowest-order, non-vanishing term is of quadrupole order:
\begin{equation}\label{quadrupole_term_SP}
\Delta V_{SP} \approx \frac{e^2 L^2 \left(1 + 3 \cos(2 \theta ) \right)}{32 \pi r^3 \epsilon_0 \epsilon_b},
\end{equation}
where $L$ is the lateral electron confinement length and $(r,\theta, \phi )$ is the location of the noise source in polar-spherical coordinates. Alternatively, for two valley states with identical envelope functions, we find that $\Delta V$ is exponentially suppressed by a factor of $e^{-d^2 k_0^2}$ to quadrupole order, where $d$ is the z-envelope width and $k_0$ is the location of the valley minimum. To evaluate Eq.~(\ref{quadrupole_term_SP}), we must estimate the typical distance $r$ between the charge trap and the qubit. To do this, we consider a slightly different system comprised of a double dot charge qubit in GaAs,
for which the energy splitting has been measured experimentally and found to be $\Delta V^{exp} \approx 1.6$~$\mu$eV.\cite{Hayashi:2003p226804}
Although this system differs from a two-electron dot, the statistics of the charge fluctuators should be similar.
 Evaluating Eq.~(\ref{deltaVdef}) for the double-dot geometry, we find
that the leading order term is a dipole contribution:
\begin{equation}\label{jkg_dd_dephasing}
\Delta V_{DD}\approx \frac{e p_0 \cos \phi \sin \theta}{4 \pi r^2 \epsilon_0 \epsilon_b },
\end{equation}
where $p_0$ is the dipole moment $e x_0$ associated with the dot separation $x_0$.
Averaging over $\theta$ and $\phi$ with $x_0 \approx 300 $~nm (the distance between the double dots considered in the experiment), we solve Eq.~(\ref{jkg_dd_dephasing}) to find $r \approx 2.9 $~$\mu$m. Inserting this into Eq.~(\ref{quadrupole_term_SP}), we obtain $\Delta V_{SP} \approx 2\times10^{-3}$~$\mu$eV for $L = 40$ nm in Si. 
Note that this value for $\Delta V_{SP}$ is likely an overestimate: if the dephasing is due to multiple charge traps (instead of the single trap we have assumed), $\Delta V_{SP}$ will be decreased. 
This is because matching to $\Delta V^{exp}$ while increasing the number of traps increases the average $r$.
Since $\Delta V_{SP} \sim1/r^3$ and $\Delta V_{exp} \sim 1/r^2$,  $\Delta V_{SP}$ decreases.
Therefore, our estimate of the $S-T$ dephasing rate for an ideal orbital first excited state is an overestimate, and may decrease due to multiple charge traps.

Now that we know the magnitude of the energy fluctuations, we can calculate the dephasing time $T_2$.  The off-diagonal elements of the density matrix decay as $e^{-\Delta \phi( \tau )}$, so the time $T_2$ is defined by $\Delta \phi( T_2 ) =1 .$ \cite{Astafiev:2004p267007} 
Following Ref.~\onlinecite{Hu:2006p100501}, the time-dependent dephasing is given by
\begin{equation}\label{defofphi}
\Delta \phi( \tau) = \frac{1}{2 \hbar^2} \int_{\omega_0}^{\infty} d \omega S (\omega) \left( \frac{\sin \omega \tau /2}{\omega/2} \right)^2,
\end{equation}
where $\omega_0$ is a low-frequency cutoff that is the inverse measurement time.
Up to this point, we have only considered the coherent evolution of the phase due to charge fluctuations. 
However, true decoherence occurs due to the statistical nature of the fluctuators. 
This effect is captured in the spectral density $S(\omega)$ of the charge noise, through the definition
\begin{equation}\label{def_of_S}
S(\omega) = \frac{1}{2 \pi} \int_{-\infty}^{\infty} d \tau e^{i \omega \tau} \left< \Delta V (\tau) \Delta V(0) \right>.
\end{equation}
As noted in Ref.~\onlinecite{Hu:2006p100501}, by examining the form of Eqs.~(\ref{defofphi}) and~(\ref{def_of_S}) we can deduce that $T_2$ for $1/f$ noise should scale as $1/\Delta V$. Thus, we can calibrate our $T_2$ to the experimental measurement via
\begin{equation}\label{t2_conversion}
T_2 \approx \left| \frac{\Delta V^{exp}}{\Delta V} \right| T_2^{exp},
\end{equation}
where $\Delta V^{exp}$ and $T_2^{exp}$ are the experimental charge qubit measurements for the energy splitting fluctuation and the dephasing time. For the double-dot charge qubit experiment referenced above, $T_2^{exp}\approx 1$~ns, which leads to $T_2^{SP}\approx 0.8$~$\mu$s for our two electron dot with an orbital-like first excited state. Thus, the dephasing rate for orbital excitations due to charge noise  in Si is $\Gamma_{ST}^{Charge} \approx 1.3$~MHz, which is on the same order as the phonon-induced dephasing in Si, and far slower than the dominant dephasing mechanism in GaAs. We plot this dephasing rate as a function of $L$ alongside the phonon dephasing mechanisms in Fig.~\ref{phonon_figure}.

This long dephasing time is due to the fact that for a perfectly harmonic confinement potential, the dipole term of $\Delta V_{SP}$ vanishes. However, in realistic systems, potential anharmonicity and interface roughness result in a non-vanishing dipole moment that can be more important than the quadrupole term in Eq.~(\ref{quadrupole_term_SP}). If our confinement potential is severely anharmonic, we expect that we would have a dipole contribution similar to Eq. (\ref{jkg_dd_dephasing}), but with a moment of  $p_0 \lesssim eL$. Further, a disordered interface might also introduce a dipole moment. 
As an example, simulations of the typical devices used in Refs.~\onlinecite{Thalakulam:2010p183104} and \onlinecite{Thalakulam:2011p045307} find $p_0/e = 1.8$~nm.

\begin{figure}[tb!] 
\begin{center} 
\includegraphics[width=1.0 \linewidth]{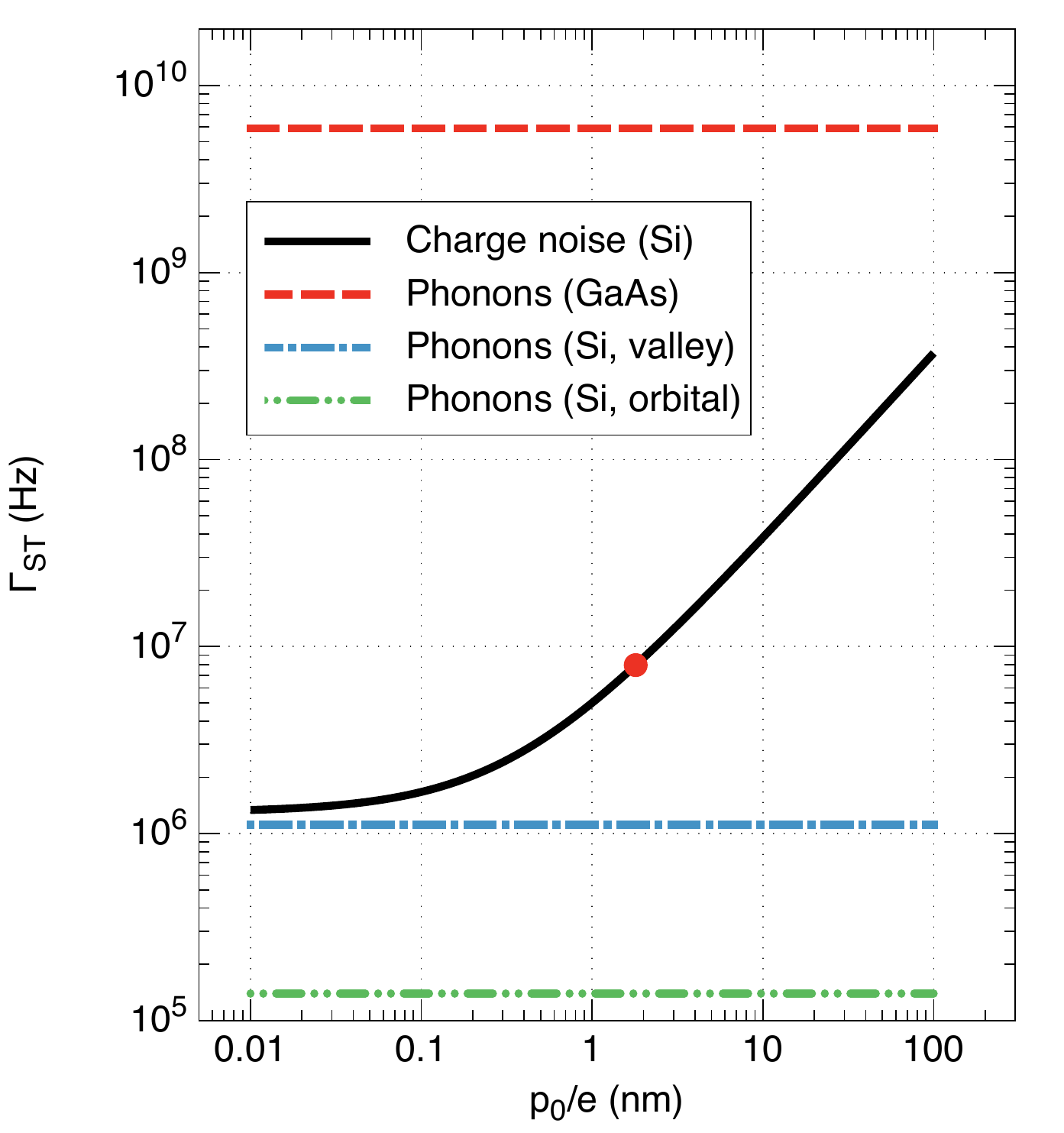}
\end{center} 
\caption{(Color online) Singlet-triplet dephasing rate $\Gamma_{ST}$ due to charge noise and electron-phonon coupling as a function of effective dipole moment $p_0$. In this plot, a constant quadrupole contribution of $1.3$~MHz (Eq.~(\ref{quadrupole_term_SP})) is added to the dipole contribution, which is the estimated dephasing rate from charge noise for a dot with a purely harmonic confinement potential, for which $p_0/2$ is zero. Here, we have set the lateral electron confinement to be $L=40$~nm, the vertical confinement to be $d=3$~nm, and have assumed an orbital excited state.
The charge noise curve for Si is estimated using a dielectric constant $\epsilon_0^{Si} = 11.7 \epsilon_{vac}$.
In a perfectly harmonic dot, $p_0 \approx 0$, but anharmonicity and disorder can introduce a dipole moment of $p_0 \lesssim eL$. In GaAs we expect phonon-mediated dephasing to be most important, but in Si quantum dots charge noise can easily dominate. The circle marker indicates the estimated dephasing due to charge noise at $p_0/e = 1.8$~nm, an estimated dipole moment for realistic devices.\cite{Thalakulam:2010p183104,Thalakulam:2011p045307}
\label{charge_noise_fig}
}
\end{figure}

Fig.~\ref{charge_noise_fig} shows the dephasing rate due to charge noise as a function of dipole moment $p_0$, obtained by using Eqs.~(\ref{jkg_dd_dephasing}) and~(\ref{t2_conversion}).
The figure also shows the dominant dephasing rates from phonons calculated in Sec.~\ref{electron_phonon_dephasing} and listed in Table~\ref{dephasing_table}, which are essentially independent of dipole moment because there is substantial electron-phonon coupling even for perfectly harmonic confinement potentials.
Within our approximations, we see that phonon-mediated dephasing is the most important mechanism in GaAs, but in Si charge noise can easily dominate.

\section{Discussion}\label{discussion}

In this paper, we addressed dephasing due to electron-phonon coupling and charge noise for two-electron states in a single quantum dot in both GaAs and Si. 
For the electron-phonon coupling, we found that in GaAs the main contribution to dephasing is due to polar coupling to optical phonons, and that the dephasing rate was of order gigahertz.
In Si, phonon-mediated dephasing rates are much lower than in GaAs because there is no polar coupling to phonons,
Intervalley processes are more important than intravalley processes, since phonons that couple valleys in silicon are extremely short-lived. 
The intervalley coupling to phonons leads a dephasing rate for silicon of order megahertz.

We found that charge noise for an orbital first excited state in a perfectly harmonic quantum dot with no disorder is strongly suppressed because the singlet-triplet energy splitting fluctuations produced by a remote perturbing potential in this case are of quadrupole order, while for a double-dot charge qubit they are of dipole order. 
As has been noted previously,\cite{Culcer:arXiv:1107.0003} pure valley states in Si are even more favorable, as they are largely immune to charge noise, in that both the dipole and quadrupole terms are suppressed by a factor of $e^{-d^2 k_0^2}$. 
However, the introduction of either anharmonicity (for orbital excitations) or disorder (for valley excitations) leads to non-vanishing dipole moments up to the order of the lateral wavefunction extent.
For either type of excited state in Si, this can become the dominant dephasing mechanism. 
Our estimated dephasing rate due to charge noise, based on calculations in typical dots, is of order $10$~MHz. This rate is fast enough to dominate the dephasing in silicon, but likely not in GaAs.

Our calculations suggest that two-electron, singlet-dot systems in Si have substantially better dephasing properties than those in GaAs. 
This is because the polar coupling for optical phonons, which mediates fast dephasing in GaAs, is absent in Si.
Within Si, to reduce the dephasing in this system, the critical figure to optimize is the effective dipole moment of the charge density difference between the first excited state and the ground state. As indicated in Fig. \ref{charge_noise_fig}, we estimate that unless this dipole moment is reduced below $p_0/e\sim 1$~nm,  charge noise is expected to be the dominant dephasing mechanism. Below that threshold, the electron-phonon coupling (for valley excitations) and quadrupole-order charge noise (for orbital excitations) are expected to be the dominant dephasing mechanisms.

\section{Acknowledgements}
The authors are grateful for useful discussions with M.A.~Eriksson, C.~Tahan, R.~Joynt, J.M.~Taylor, and H.J.~Maris.
JKG gratefully acknowledges support from the NSF. MF and SNC acknowledge support by NSA/LPS through ARO (W911NF-08-1-0482) and by NSF (PHY-1104660). XH acknowledges support by  NSA/LPS through ARO (W911NF-09-1-0393) and by NSF  (PHY-1104672).

\bibliography{siliconQCgamble}

\end{document}